\pdfoutput=1

\documentclass[sn-mathphys, Numbered, iicol]{sn-jnl}

\usepackage{graphicx}%
\usepackage{multirow}%
\usepackage{amsmath,amssymb,amsfonts}%
\usepackage{amsthm}%
\usepackage{mathrsfs}%
\usepackage[title]{appendix}%
\usepackage{xcolor}%
\usepackage{textcomp}%
\usepackage{manyfoot}%
\usepackage{booktabs}%
\usepackage{algorithm}%
\usepackage{algorithmicx}%
\usepackage{algpseudocode}%
\usepackage{listings}%
\usepackage{cleveref}
\usepackage{ulem}
\usepackage{xspace}
\usepackage{mhchem}
\usepackage{subcaption}
\makeatletter
\newcommand*{\myfnsymbolsingle}[1]{%
  \ensuremath{%
    \ifcase#1
    \or 
      \S%
    \or 
      \dagger
    \or 
      a
    \or 
      b
    \or 
      c
    \or 
      d
    \else 
      \@ctrerr  
    \fi
  }%
}   
\makeatother

\usepackage{alphalph}
\newalphalph{\myfnsymbolmult}[mult]{\myfnsymbolsingle}{}

\usepackage{fnpct} 

\theoremstyle{thmstyleone}%
\theoremstyle{thmstyletwo}%

\theoremstyle{thmstylethree}%

\Crefname{figure}{\text{Fig.}}{\text{Figs.}}
\Crefname{section}{\text{Section}}{\text{Sections}}

\newcommand{\pab}[2]{\frac{\partial #1}{\partial #2}}
\newcommand{\ppabc}[3]{\frac{\partial^2 #1}{\partial #2\partial #3}}

\newcommand{\fl}{Forward Laplacian\xspace}
\newcommand{\namett}{\texttt{LapNet}\xspace}
\newcommand{\name}{LapNet\xspace}

\newcommand{\sptable}[1]{Supplementary Table #1}

\usepackage{amsmath,amsfonts,bm}

\def\eqref#1{equation~\ref{#1}}

\def\1{\bm{1}}

\def\rvg{{\mathbf{g}}}
\def\rvh{{\mathbf{h}}}

\def\rvr{{\mathbf{r}}}

\def\rvv{{\mathbf{v}}}

\def\rvx{{\mathbf{x}}}
\def\rvy{{\mathbf{y}}}
\def\rvz{{\mathbf{z}}}

\DeclareMathAlphabet{\mathsfit}{\encodingdefault}{\sfdefault}{m}{sl}
\SetMathAlphabet{\mathsfit}{bold}{\encodingdefault}{\sfdefault}{bx}{n}

\begin{document}

\title[Article Title]{Forward Laplacian: A New Computational Framework for Neural Network-based Variational Monte Carlo}


\author[1,2]{\fnm{Ruichen} \sur{Li}}\nomail\intern{Interns at ByteDance Research}\equalcont{These authors contributed equally to this work.}
\author[1]{\fnm{Haotian} \sur{Ye}}\nomail\equalcont{These authors contributed equally to this work.}
\author[1,2]{\fnm{Du} \sur{Jiang}}\nomail\intern{Interns at ByteDance Research}
\author[2]{\fnm{Xuelan} \sur{Wen}}\nomail
\author[1]{\fnm{Chuwei} \sur{Wang}}\nomail
\author[2]{\fnm{Zhe} \sur{Li}}\nomail
\author[2]{\fnm{Xiang} \sur{Li}}\nomail
\author[1,a]{\fnm{Di} \sur{He}}\email{dihe@pku.edu.cn}
\author[1,b]{\fnm{Ji} \sur{Chen}}\email{ji.chen@pku.edu.cn}
\author[2,c]{\fnm{Weiluo} \sur{Ren}}\email{renweiluo@bytedance.com}
\author[1,d]{\fnm{Liwei} \sur{Wang}}\email{wanglw@cis.pku.edu.cn}



\affil[1]{\orgname{Peking University}}

\affil[2]{\orgname{ByteDance Research}}



\abstract{

Neural network-based variational Monte Carlo (NN-VMC) has emerged as a promising cutting-edge technique of \textit{ab initio} quantum chemistry. However, the high computational cost of existing approaches hinders their applications in realistic chemistry problems. Here, we report the development of a new NN-VMC method that achieves a remarkable speed-up by more than one order of magnitude, thereby greatly extending the applicability of NN-VMC to larger systems. Our key design is a novel computational framework named \fl, which computes the Laplacian associated with neural networks, the bottleneck of NN-VMC, through an efficient forward propagation process. We then demonstrate that \fl is not only versatile but also facilitates more developments of acceleration methods across various aspects, including optimization for sparse derivative matrix and efficient neural network design. Empirically, our approach enables NN-VMC to investigate a broader range of atoms, molecules and chemical reactions for the first time, providing valuable references to other \textit{ab initio} methods. The results demonstrate a great potential in applying deep learning methods to solve general quantum mechanical problems.
}

\maketitle

\small

\section{Main}\label{sec: Main}
\stepcounter{footnote}\footnotetext{These authors carried out this work as interns at ByteDance. }
\stepcounter{footnote}\footnotetext{Equal Contribution}
\stepcounter{footnote}\footnotetext{dihe@pku.edu.cn}
\stepcounter{footnote}\footnotetext{ji.chen@pku.edu.cn}
\stepcounter{footnote}\footnotetext{renweiluo@bytedance.com}
\stepcounter{footnote}\footnotetext{wanglw@pku.edu.cn}

Accurately solving the time-independent electronic Schr\"{o}dinger equation can yield the fundamental properties of a given quantum mechanical system. Quantum Monte Carlo (QMC) \cite{needs_variational_2020, foulkes_quantum_2001} is one of the most important \textit{ab initio} methods for solving the Schrödinger equation, widely employed in various scenarios in quantum chemistry. However, in QMC, the accuracy of the solution heavily relies on the choice of the ansatz, which requires significant expertise. This limitation restricts the adaptability of QMC compared to other deterministic approaches, such as CCSD(T), which is widely considered as the ``gold standard".

Recently, deep learning has revolutionized the field of quantum chemistry for obtaining more accurate solutions to the Schrödinger equation. One of the pioneering approaches is Neural Network-based Variational Monte Carlo (NN-VMC), such as FermiNet \cite{ferminet,spencer_better_2020} and PauliNet \cite{hermann2020deep}. The ground state wavefunction can be obtained through the variational principle, by which the expectation value of the energy is minimized. Based on this fact, NN-VMC methods utilize deep neural networks to parameterize the wavefunction and optimize the network parameters with the energy serving as the loss function. Benefiting from the remarkable capacity of neural networks, NN-VMC methods demonstrated promising results to attain chemical accuracy for diverse systems. However, it is essential to note that these methods often incur significant computational costs during model training. For instance, learning a wavefunction for the benzene dimer system requires approximately 10,000 GPU hours \cite{glehn2023a} on modern hardware, making it challenging to scale up toward larger systems.

In this work, we tackle the challenge of computational efficiency in NN-VMC methods, especially for large-scale systems. Note that NN-VMC needs to compute the Laplacian with respect to the neural network input to obtain the loss. Through our investigation, the calculation of this term consumes a substantial proportion of the overall training time and becomes the primary bottleneck in the learning procedure. In detail, previous works first compute the Hessian matrix using the auto differentiation (\texttt{AutoDiff}) method in deep learning toolkits and then derive the Laplacian by taking the trace. Such a process requires performing forward propagation and backward propagation multiple times, significantly slowing down the overall training speed of NN-VMC. See Section \ref{subsec: method forward lap} for a detailed analysis.

Observing this problem, we first develop a novel computational framework named \fl. In contrast to the commonly used method that indirectly derives the Laplacian from the Hessian, \fl directly computes the value through a carefully crafted forward propagation process, which we mathematically show is much more efficient as it eliminates unnecessary computation and propagation. Second, we demonstrate that this way of computation not only accelerates the process but also paves the way for developing advanced techniques in NN-VMC. In particular, we illustrate that in the \fl associated with NN-VMC ansatz, many intermediate derivatives exhibit sparsity and can be optimized to a considerable extent. We also design an efficient neural network architecture called \name, which can better exploit the strength of the \fl method using carefully designed attention blocks with sparse derivatives. Together, these developments allow us to study atoms, molecules, and chemical reactions beyond the capability of existing NN-VMC packages.

In the following sections, we first evaluate our method on a wide range of systems in terms of calculating the absolute energy following  \cite{glehn2023a,ferminet,spencer_better_2020}. All results consistently indicate that \name, coupled with the \fl method, obtains accurate energy estimation while significantly reducing the computational cost of model training. Given these promising results, we further explore whether our approach can learn a more useful quantity---the relative energy---across different practical scenarios, including the barrier of chemical reactions, the ionization energy of transition metals, and the noncovalent interaction between molecules. The results demonstrate, for the first time, that the relative energies obtained using NN-VMC methods align with those obtained using gold-standard computational methods and experimental results, suggesting a great potential of using deep learning to solve quantum mechanical systems. 

\section{Results}\label{sec: Results}

\begin{figure*}[t]
    \centering
    \includegraphics[width=0.95\textwidth]{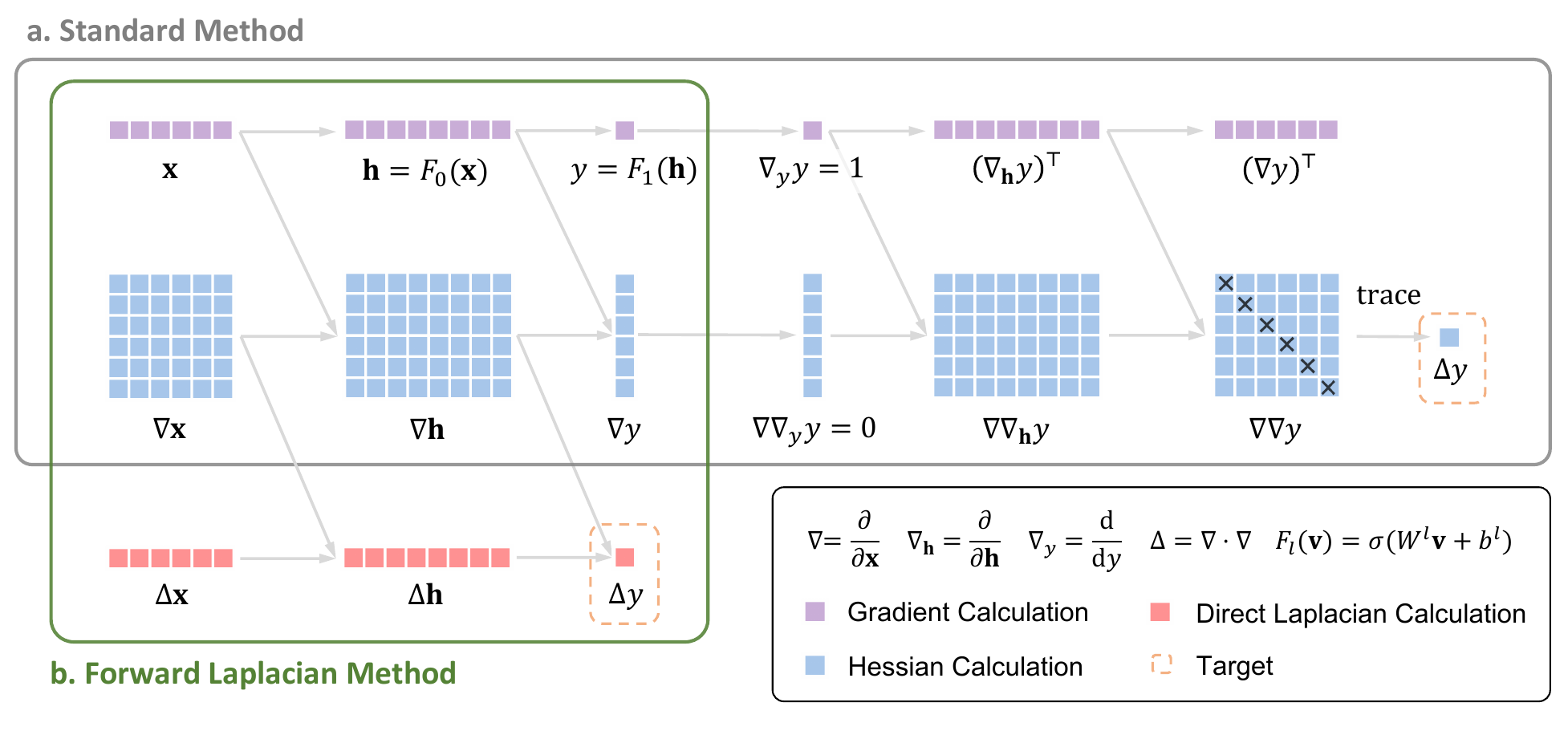}
    \caption{\small Illustration of the computation process for Laplacian with the standard method and the proposed \fl method. The standard method is presented in the grey box, and our proposed method is presented in the green box. From the figure, we can see that the standard method requires the calculation of Hessian in deriving Laplacian, which is unnecessary and inefficient. In contrast, our method obtains the Laplacian through a carefully crafted forward propagation process, and thus approximately halves the total computational cost.}
    \label{fig:FL-vs-Hess}
\end{figure*}

\subsection{\fl Framework}
\label{2.1}
The most time-consuming part of NN-VMC is the calculation of the Laplacian, i.e., the divergence of the gradient of the network's output with respect to the coordinates of electrons. In modern deep learning toolkits, one of the most commonly used methods for obtaining the Laplacian is to first compute the Hessian matrix with the \texttt{AutoDiff} package and then take its trace \cite{ferminet,spencer_better_2020,hermann2020deep,glehn2023a}. However, deriving the Hessian matrix to compute the Laplacian is unnecessary and inefficient, making it far from ideal. We develop a brand-new computational framework, named \texttt{\fl}, that directly calculates the Laplacian and can significantly improve the computation efficiency of NN-VMC. 
For the sake of simplicity and without loss of generality, we introduce the implementation and the advantage of \fl on the Multi-Layer Perceptron (MLP) network. A detailed analysis of the \fl method for general neural networks will be discussed in Section \ref{sec: Methods}.

Let $\rvx$ denote the input variable (e.g., the electron coordinates) and $\phi$ denote an arbitrary MLP model. Usually, $\phi$ can be decomposed into a sequence of linear transformations and non-linear activation functions, i.e., $\phi(x)={F}_L\circ {F}_{L-1}\circ ...\circ{F}_{0}(x)$, where $F_{l}(\rvh)=\sigma(W^{l}\rvh+b^{l})$, $l=0,\cdots, L$. $W^{l}$ and $b^{l}$ are learnable parameters and $\sigma$ is an element-wise activation function. For ease of reference, we denote $\rvh^{l+1}={F}_{l}(\rvh^{l})$ as the hidden value and ${y}={F}_{L}(\rvh^L)$ as the final output. 
We denote the derivative with respect to variable $\rvh$ as $\nabla_{\rvh}$. For simplicity, we use the notation $\nabla$ for the derivative with respect to input and $\Delta$ for the Laplacian with respect to input. For a vector-valued function, its Laplacian is derived through applying $\Delta$ to each element. To compute $\Delta\phi(\rvx)$, previous method first computes the first-order derivative $\nabla \phi$ through a forward propagation followed by a backward propagation:
\begin{equation}
\begin{aligned}
   & {\bf x} \rightarrow \cdots \rightarrow {\bf h}^{l} \rightarrow {\bf h}^{l+1} \rightarrow \cdots \rightarrow  y \\
    &\quad \rightarrow \nabla_{{\bf h}^{L}}y \rightarrow  \cdots \rightarrow  \nabla_{{\bf h}^{l}}y \cdots \rightarrow  \nabla_{\bf x} y ,
\end{aligned}
 \end{equation}
where the chain rule in deriving $\nabla_{{\bf h}^{l}}y$ is given by $\nabla_{\rvh^l} y = \nabla_{\rvh^{l}}F_l(\rvh^{l})\nabla_{\rvh^{l+1}} y$. The Hessian matrix $\nabla^2\phi$ is calculated through another pass using the following chain rule:
\begin{align}
    \nabla \rvh^{l+1} &= \nabla \rvh^{l}\nabla_{\rvh^{l}} F_l(\rvh^{l}) \label{hessian forward:mlp grad},\\
         \nabla\nabla_{\rvh^l}y &=\nabla\nabla_{\rvh^{l+1}} y\ \nabla_{\rvh^l}F_l(\rvh^{l})^{\top} \nonumber \\
          &\quad + \ [(\nabla\rvh^l \cdot \nabla_{\rvh^l})\nabla_{\rvh^l}F_l(\rvh^{l})]\nabla_{\rvh^{l+1}} y\ . 
         \label{hessian backward:mlp gradient}
\end{align}
The Laplacian is then obtained by taking the trace and the whole computation graph of the method is visualized in Fig. \ref{fig:FL-vs-Hess}a. 

Unlike the previous method, we demonstrate that the Laplacian can be calculated efficiently in a direct way. In the forward pass, in each layer $l$, besides calculating the value of hidden state $\rvh^l$, we additionally calculate two terms: the first-order derivative $\nabla \rvh^l$, as well as the intermediate Laplacian term $\Delta\rvh^l$. 
\begin{equation}       
\label{eq: tuple propagation}
\left[                
  \begin{array}{c}   
    {\bf x} \\  
    {\nabla \rvx} \\
    {\Delta\rvx}\\
  \end{array}
\right]  \rightarrow \cdots \rightarrow
\left[                
  \begin{array}{c}   
    {\rvh^l} \\  
    {\nabla \rvh^l} \\
    {\Delta\rvh^l}\\
  \end{array}
\right]\rightarrow \cdots \rightarrow
\left[                
  \begin{array}{c}   
    {y} \\  
    {\nabla y} \\
    {\Delta y}\\
  \end{array}
\right].
\end{equation}
The key insight in our method is that the calculation of all three terms in layer $l+1$ only requires the term values in layer $l$, where the chain rule is given by 
\begin{align}
    \label{forward:func}
    \rvh^{l+1} &= {F}_l(\rvh^{l}) \\
    \label{mlp grad foward FL}
    \nabla \rvh^{l+1} &=\nabla \rvh^{l} \nabla_{\rvh^{l}} F_l(\rvh^{l}) , \\ 
    \Delta \rvh^{l+1} &= (\nabla_{\rvh^l} F_{l}(\rvh^{l}))^\top\ \Delta \rvh^{l}  \nonumber\\ 
    \label{lap mlp forward tr}
    &\quad + \text{tr}((\nabla \rvh^{l})^\top\nabla \rvh^{l}\nabla^2_{\rvh^{l}} F_{l}(\rvh^{l}) ).
\end{align}
In such a way, all terms are computed iteratively from the first layer, and the value of $\Delta\phi(\rvx)$ will be obtained in the final layer. This method calculates the Laplacian in a single forward pass without computing $\nabla^2\phi$, thereby avoiding redundant computations. Additionally, the memory requirements of \fl is not significant as the memory that stores intermediate tuple in eq. (\ref{eq: tuple propagation}) can be released in the succeeding forward process. The whole computation graph of the proposed method is visualized in Fig. \ref{fig:FL-vs-Hess}b. In Section \ref{sec: Methods}, we provide a detailed analysis of the computational complexity of the \fl method, demonstrating that the cost can be approximately halved compared to the previous method. 
It's important to note that the efficiency is not attributed to any approximation that loses precision but rather to an alternative computational strategy. Hence, our method serves as a more efficient replacement for the previous method in the general setting.

\begin{figure*}[t]
    \centering
    \includegraphics[width=1.0\textwidth]{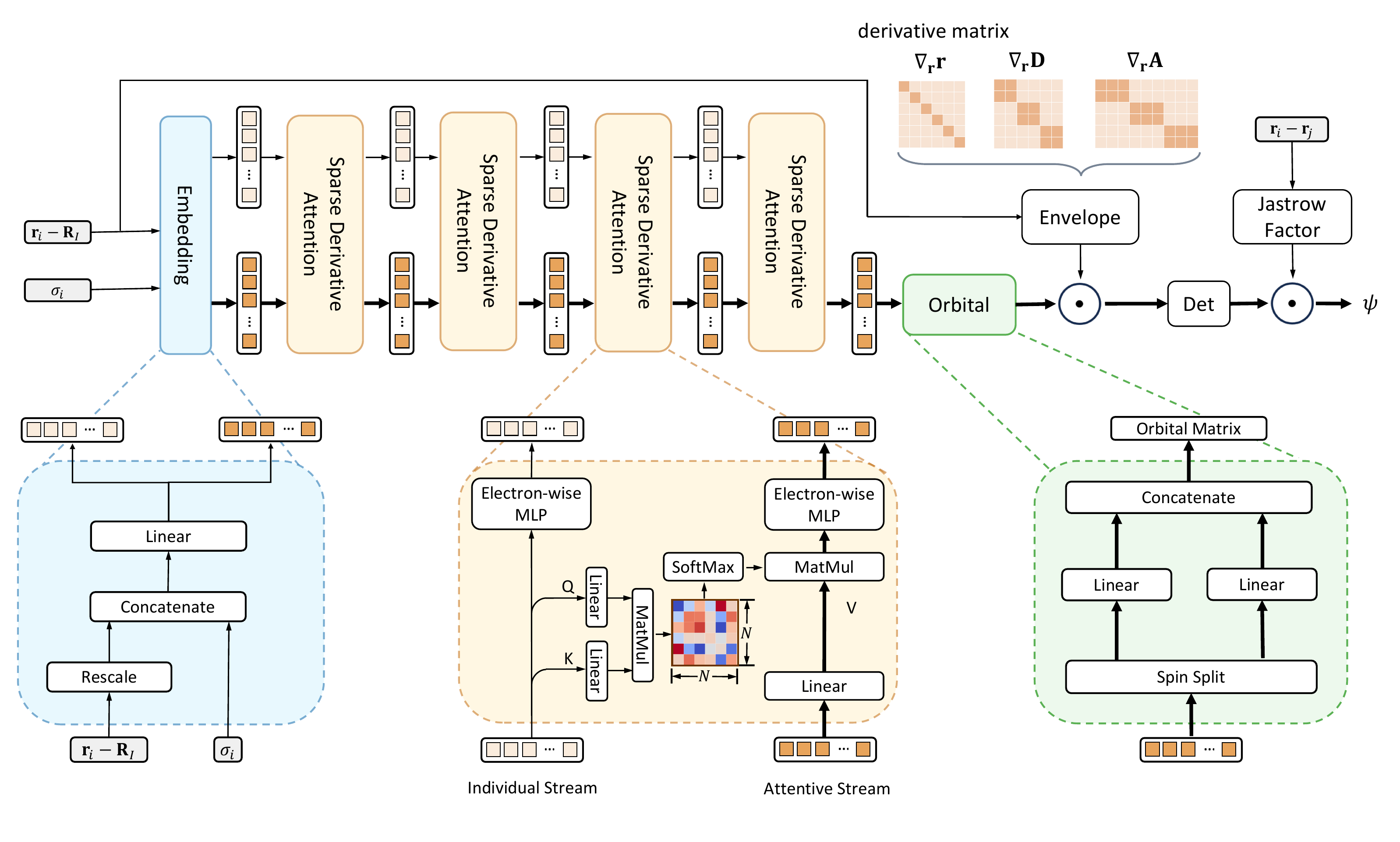}
    \caption{\small {\bf The \name Architecture.} The blue block represents the embedding layer, the yellow block represents the sparse derivative attention block, and the green block represents the orbital mapping. The top right part illustrates the derivative sparsity in the envelope function. $\rvr$ is the input electron position, $\mathbf D, \mathbf A$ are intermediate matrices. The dark color represents non-zero elements and the light color represents zero elements. The derivative matrices are structurally sparse, thus we can reduce the cost of storing and operating these matrices. A detailed explanation of each block will be shown in Section \ref{sec: ansatz}. }
    \label{fig:network}
\end{figure*}

While the \fl method offers superior efficiency from a computational point of view, it is crucial yet challenging to integrate it into modern deep learning toolkits and make it compatible with existing frameworks, as the entire training algorithm still heavily relies on many other \texttt{AutoDiff} functionalities, e.g., the backward gradient calculation for the network parameters. We implement the \fl method in \texttt{JAX} \cite{jax2018github}, a greatly flexible and efficient open-sourced toolkit developed by Google. We carefully overload all the related functions such that when the input of $f(\rvx)$ is a tuple $\left(\rvx,\nabla \rvx,\Delta \rvx\right)$, the output will automatically become $\left(f(\rvx),\nabla f(\rvx), \Delta f(\rvx)\right)$. Additionally, for a wide range of elementary operations used in the NN-VMC methods, we manually define their forward propagation rules (e.g., eq. (\ref{lap mlp forward tr})) and optimized their efficiency.

\subsection{Accelerating NN-VMC using the Forward Laplacian Framework} \label{sec: sparsity}

In the previous subsection, we introduce a new computational framework that can accelerate the calculation of Laplacian in general settings. In this subsection, we demonstrate that using the proposed framework, we can achieve an even more substantial acceleration rate for some specific classes of neural networks, such as NN-VMC ansatz. 

One of the key designs in Forward Laplacian is using forward propagation only. When we applied \fl to an NN-VMC ansatz, we found that many intermediate outputs along the propagation naturally exhibit sparsity and can be further optimized to a considerable extent. 
For example,  the \textit{envelope function} (refer to Fig. \ref{fig:network}) 
 of any NN-VMC ansatz guarantees that the wavefunction decays exponentially as electrons move away from atoms. 
It sequentially computes the relative distance matrix $\mathbf D$ between electrons and nuclei, and the exponentially decaying feature matrix $\mathbf A$. Mathematically, $\nabla_\rvr \rvr, \nabla_\rvr \mathbf D, \nabla_\rvr \mathbf A$ are all block-diagonal, where $\rvr$ denotes the electron positions. Matrix operations can be optimized by leveraging this sparse property. We optimize the \fl method to preserve only the non-zero elements and remove others, resulting in an $\mathcal O(N)$ efficiency gain in both memory and computation, where $N$ is the number of electrons. The improvement can be applied to various components beyond the envelope function, including the Hartree-Fock function and the Jastrow factor \cite{PhysRevB.70.235119}. Remarkably, this technique is only applicable to our framework, as the intermediate variables (e.g., $\nabla\nabla_{\rvh^l}y$) in the standard approach are not sparse.

\begin{figure*}[htbp]
    \includegraphics[width=\textwidth]{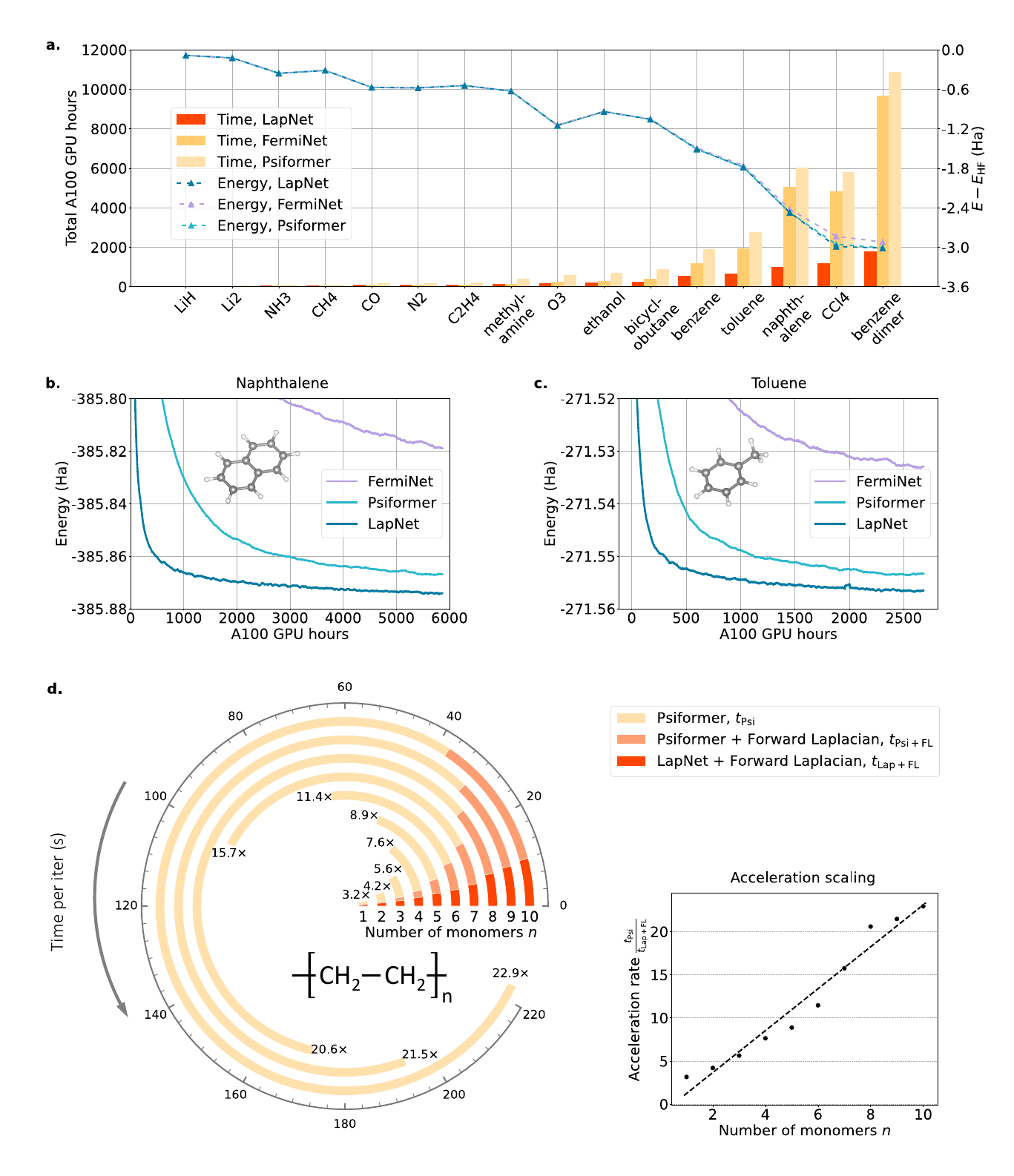}
    \centering
    \caption{\small {\bf Efficiency and performance comparison of different NN-VMC methods.} {\bf a}, Total training time and energy of \name, FermiNet, and Psiformer for 200,000 iterations.
    The FermiNet and Psiformer results are from \cite{glehn2023a}. See \sptable{6} for a detailed energy comparison.
    {\bf b}, {\bf c}, \name's training curve on naphthalene and toluene, compared against the curve of Psiformer and FermiNet.
    {\bf d}, 
    To investigate the scaling effect of the computational cost with system size, we measure the per-iteration time of different models on Polyethylene systems with varying molecule numbers. We study three models for each system, Psiformer, Psiformer with Forward Laplacian, and \name. The speed-up rate of LapNet compared to the original Psiformer is presented at the endpoint of each bar. We also illustrate the dependency between the speed-up rate and the system size on the right side, and the results suggest an almost linear relationship between the acceleration rate and the size of the system.}
    \label{fig:abs-exp}
\end{figure*}

We then design a new neural network ansatz, namely the \namett, which increases the sparsity during the forward process in \fl while maintaining excellent performance compared to other NN-VMC architectures. We illustrate the structure of \name in Fig. (\ref{fig:network}). \name incorporates general attention modules, a popular component used in numerous powerful models such as GPT \cite{brown2020language} and AlphaFold2 \cite{AlphaFold2021}. We develop a \textit{sparse derivative attention} (SDA) block, which is functionally similar to the standard attention block while inherently introducing additional sparsity to the derivative matrices, thereby achieving better efficiency through the \fl method. Extensive evaluations in Section \ref{sec:2.3} demonstrate the efficiency and performance of \name.

\subsection{Efficiency and Performance}
\label{sec:2.3}
In this subsection, we demonstrate that the combination of Forward Laplacian and LapNet achieves state-of-the-art energy results while reducing the computational cost by up to 20 times.
We first benchmark \name against FermiNet \cite{ferminet} and the recently introduced Psiformer \cite{glehn2023a}, an attention-based neural network ansatz that has shown state-of-the-art performance. The architecture hyperparameters can be found in \sptable{1}.
%
%
Second, we study the runtime scaling law of different methods on polyethylene systems \ce{(C2H4)_n} over different $n$. Throughout this section, we evaluate the performance of trained models using absolute energy. 
%
For baseline approaches, we directly use the open-sourced implementation \cite{FermiNet2020github} to measure the corresponding runtime.
%
%
%
More details on network training and energy evaluation are discussed in Section \ref{sec: Methods}.

We first calculate the absolute energy on $16$ molecular systems with electron numbers ranging from $4$ to $84$ with 200,000 training iterations. 
The total training time and energy results of all methods are presented in Fig. \ref{fig:abs-exp}a. 
Both \name and Psiformer consistently produce lower, namely better, absolute energy than FermiNet across all systems. 
When compared to Psiformer, the energy results of \name either align closely with those of Psiformer within chemical accuracy, or outperform them in the case of certain larger molecules such as \ce{CCl4}.
This suggests that \name has state-of-the-art capacity for approximating the ground state wavefunction. 
Regarding the total training time, it is clear from Fig. \ref{fig:abs-exp}a that \name incurs significantly lower costs compared to the baselines.
We provide a detailed examination of training curves on naphthalene and toluene in Fig. \ref{fig:abs-exp}b and Fig. \ref{fig:abs-exp}c.
In both figures, the $x$-axis indicates the training time instead of the number of iterations, and the $y$-axis indicates the estimated energy. 
It is evident that our \name converges faster and achieves better accuracy compared to the baselines. 
This is because \name can run for more iterations within the same training period, made possible by the efficient Laplacian calculation in Forward Laplacian, leading to improved accuracy.


%
To better understand the efficiency enhancements, we investigate the per-iteration computational cost of different models using the polyethylene systems \ce{(C2H4)_n} as an example and set $n$ ranging from $1$ to $10$. We mainly study three models, one LapNet and two Psiformers. For one of the Psiformers, we use the Forward Laplacian method with sparse derivative matrix optimization, and for the other one, we use the original implementation. The computational cost and speed-up rate of all the models are shown in Fig. \ref{fig:abs-exp}d.
%
%
%
It can be seen from the figure, the \fl method can accelerate the Psiformer by 2.6 times when $n = 1$, and 5.9 times when $n$ increases to $10$. This observation is consistent with our analysis in Section \ref{sec: sparsity} that when the system size grows, the speed-up rate of using the \fl method will be more significant. 
The combination of \fl and \name generates a speed-up rate exceeding $20$ when compared to Psiformer without \fl, with $n=10$. 
This demonstrates the significant potential of our framework for larger systems. 
In addition, when comparing the computational cost of Psiformer and \name, with \fl applied to both, the speed-up rate of \name is $2$ to $3$, which showcases the efficiency improvement of the proposed sparse derivative attention block. 

\begin{figure*}[htbp]
    \includegraphics[width=0.95\textwidth]{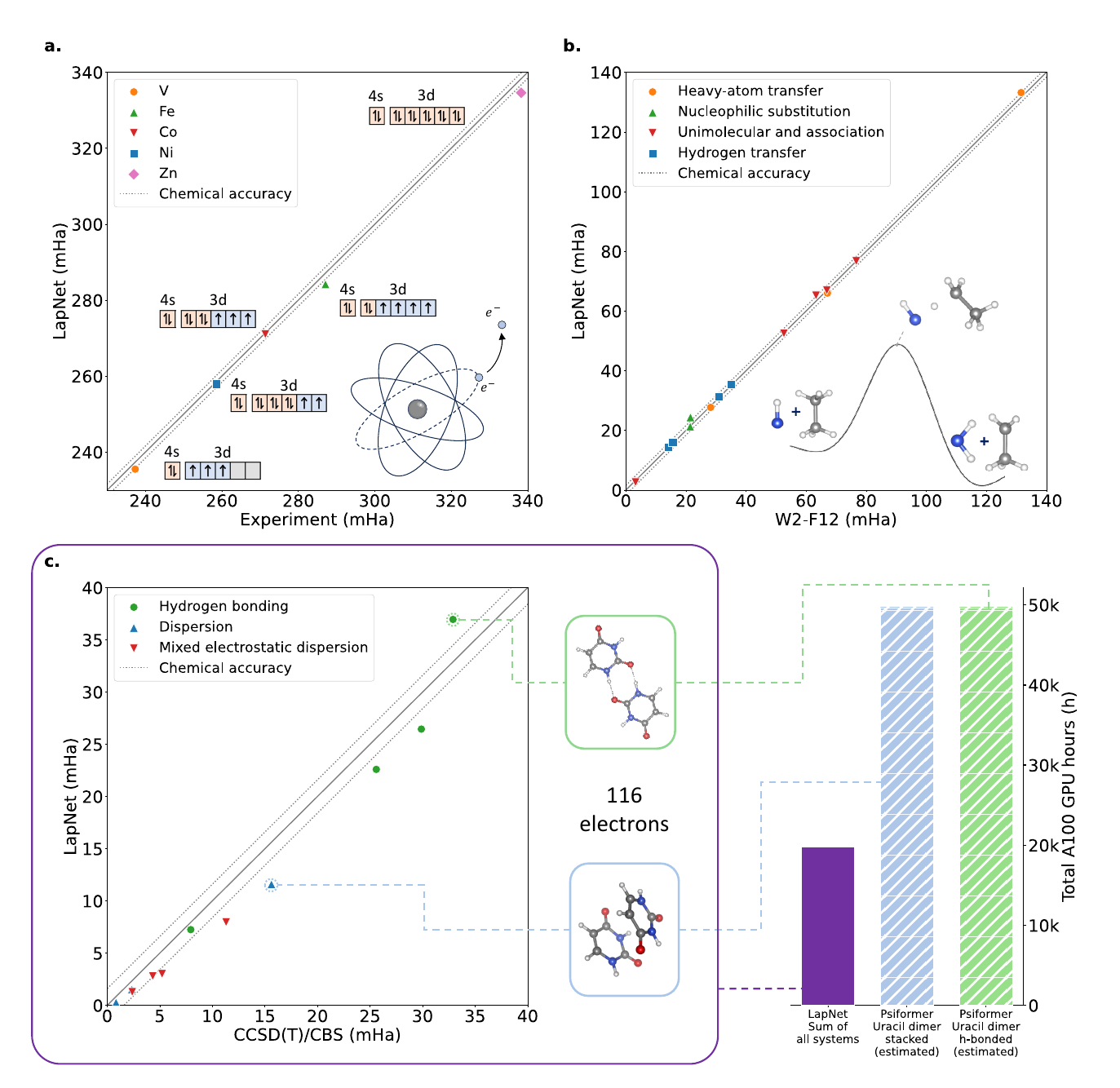}
    \centering
    \vspace{0.1cm}
    \caption{\small {\bf Relative energy estimation using \name.} 
    {\bf a}, Estimated ionization potentials of 5 third-row transition metals, compared against experimental results \cite{10.1063/1.2335444}. We also show the 4s and 3d orbital occupations of our calculated ground state next to the symbol for each atom. {\bf b}, Estimated barrier heights of 3 heavy-atom transfer, 2 nucleophilic substitution, 5 unimolecular association, and 4 hydrogen transfer reactions are plotted (see \sptable{4} for details), and W2-F12 data from \cite{C7CP04913G} are used as reference. {\bf c},  Left: Interaction energies of 4 hydrogen-bonding interaction systems, 3 dispersion interaction systems, and 4 mixed electrostatic dispersion interaction systems (see \sptable{5} for details). We take CCSD(T)/CBS results from \cite{10.1063/1.3659142} for comparison. Right: Comparison between the total A100 GPU hours of training all the systems using our method and the estimated A100 GPU hours of training 2 largest systems using Psiformer. All the energy data is given in the \sptable{7, 8, 9}. 
    } 
    \label{fig:rel-exp}
\end{figure*}

\subsection{Pushing the Limits of Relative Energy}
In quantum chemistry, relative energy holds greater significance than absolute energy in addressing practical problems. For example, the ionization energy reflects the electron affinity of molecules, the barrier height controls the kinetics of chemical reactions, and the binding energy determines the stability of chemical complexes, to name just a few.
Unfortunately, due to the huge computational cost, previous NN-VMC research mainly focused on the absolute energy of small systems, resulting in a lack of comprehensive evaluation pertaining to relative energy. Obtaining accurate relative energy is especially difficult, as the energy scale for relative energy is typically hundreds or thousands of times smaller than that of absolute energy.
It is worth noting that although examples of relative energy calculations were reported and discussed in several NN-VMC studies, accurate calculations were only possible for rather small systems and extrapolation techniques were further employed to reduce the error \cite{Ren2023,paulinet_excited}.
In this paper, we comprehensively evaluate the relative energy in more challenging and practical systems, thanks to the efficiency improvement brought by \fl and \name.  
Specifically, we calculate the ionization potential of $5$ metal atoms, reaction barrier heights of $14$ reactions, and noncovalent dissociation energy for $11$ systems. We compare \name with established benchmarks, aiming to better understand the strengths and limitations of the NN-VMC framework. 

The ionization potential of an atom 
can be accurately measured \cite{osti_6008300,Page:90,Sohl:90,10.1063/1.468462} through experiments and are widely used to benchmark the \textit{ab initio} calculation methods. We run \name to evaluate the ionization potential of five third-row transition metal atoms, namely vanadium (\ce{V}), iron (\ce{Fe}), cobalt (\ce{Co}), nickel (\ce{Ni}), and zinc (\ce{Zn}). We train \name separately on atoms and the corresponding ions, compute the energy difference as the reported ionization potentials, and compare with the experimental value where the relativistic effects are removed. As shown in Fig. \ref{fig:rel-exp}a, \name can accurately estimate the ionization potentials and achieve chemical accuracy for three out of five systems (\ce{V, Co, Ni}), and the errors on \ce{Fe, Zn} are relatively larger ($2.9$ mHa and $3.7$ mHa, respectively). 
To further study the electronic structure of these transition metals, we calculate the orbital occupations of our wavefunction in outer $4s$ and $3d$ shells, as shown in Fig. \ref{fig:rel-exp}a.
The calculated orbital occupations align perfectly with experimental observation, illustrating that our LapNet-based wavefunction is capable of not only producing accurate energy results but also encoding the correct information about the electronic structure.



The barrier height represents the energy difference between the transition state and reactants, determining the minimum energy to activate a chemical reaction and the affecting reaction rate. Therefore, accurate prediction of the barrier is crucial
for catalyst design \cite{spiekermann_high_2022}. However, popular \textit{ab initio} methods such as density functional theory (DFT) and CCSD(T) struggle to accurately describe transition states because chemical bonds are partially broken and formed during chemical reactions. In this regard, we have chosen 14 distinct barrier heights from the BH76 dataset \cite{C7CP04913G} encompassing various reaction types such as hydrogen transfer, heavy atom transfer, nucleophilic substitution, and unimolecular association reactions. We use the W2-F12 level of theory \cite{10.1063/1.3697678} in \cite{C7CP04913G} as references. As shown in Fig. \ref{fig:rel-exp}b, \name achieves chemical accuracy for $12$ out of $14$ reactions, where the average difference is only $0.7$ mHa. These results demonstrate that \name can accurately estimate the barrier heights.

Non-covalent interactions are critical for maintaining the structure of large chemical and biological molecules. 
Due to the relatively large contribution of dynamic correlations, standard quantum chemistry methods may face difficulties \cite{al2021interactions}.  
To further demonstrate the powerfulness of \name, we train \name on $11$ systems selected from the S22 noncovalent interaction dataset \cite{jurevcka2006benchmark}, where the largest contains $116$ electrons. The selected data encompasses all types of noncovalent interactions in S22, including hydrogen bonding interactions, dispersion interactions, and mixed
electrostatic dispersion interactions. To estimate the binding energy of a noncovalent system, we follow the calculation setting in \cite{Ren2023} and calculate the energy difference between the dissociated state and the equilibrium state. As shown in Fig. \ref{fig:rel-exp}c, the average difference over $11$ systems is within $2.2$ mHa compared to the CCSD(T) benchmark, and \name achieves chemical accuracy for $5$ out of $11$ systems. 
Our superior efficiency allows for the fine-grained investigation of large systems with more than a hundred electrons and sheds light on future improvement of NN-VMC methods. For instance, across the analysis of 11 systems, we notice that \name typically overestimates the energies of equilibrium states since their electron structures are more complex than that of dissociated states. Inspired by this observation, future NN-VMC methods should focus more on expressing the electron structures of equilibrium states to achieve better calculation results.

We would like to highlight that this comprehensive investigation into relative energies would not be possible without the help of \fl and \name. Prior to our work, the largest molecular system analyzed by previous NN-VMC methods was the benzene dimer system with 84 electrons. However, their training requires more than $10,000$ A100 GPU hours, while ours requires only $1,800$ GPU hours, giving a $6$ times acceleration. In addition, \name is able to estimate even larger systems, such as the uracil dimer with $116$ electrons. On the right side of Fig. \ref{fig:rel-exp}c, we present the GPU hours to compute all the 11 systems with \name and compare it with the computational cost of Psiformer to finish the two largest systems, uracil dimer stack and uracil dimer h-bonded. Due to the huge computational cost, we only estimate the GPU hours of these two training process using \texttt{per\_iteration\_cost} $\times$ \texttt{num\_iteration}, where \texttt{num\_iteration} is set to 200,000, following all previous works \cite{ferminet,glehn2023a}. It can be clearly seen that calculating all the systems using our method is even faster than calculating one system using the original Psiformer.

In summary, we demonstrate the potential of \name to estimate different types of relative energies. The significantly improved efficiency allows a more comprehensive study of a wide range of large chemical systems, which is beneficial for the development of NN-VMC methods and the field of computational quantum chemistry.

\section{Discussion}\label{sec: Discussion}
NN-VMC methods have been used for ground-state energy estimation in a wide range of applications, including molecular systems \cite{ferminet, hermann2020deep, spencer_better_2020, han_solving_2019,lin2023explicitly,abrahamsen2022taming,gerard2022goldstandard,glehn2023a}, solid systems \cite{li_ab_2022, PhysRevResearch.4.023138}, and homogeneous electron gas  \cite{wilson2022wave, PhysRevLett.130.036401, pescia2023message}. The approach has been extended to compute other important quantities, such as  excitation energy \cite{paulinet_excited},  inter-atomic force \cite{qian2022interatomic} and electric polarization \cite{li2023electric}. Recently, several works further combined NN-VMC methods with other classical methods, such as Effective Core Potential \cite{PhysRevResearch.4.013021} and Diffusion Monte Carlo \cite{Ren2023,dmcGao}.  Refs. \cite{deeperwin,gao2022abinitio,gao2023samplingfree,gao2023generalizing,scherbela2023towards} learn universal representations of wavefunctions on multiple systems to enable transfer learning to unseen system configurations. Notably, in all of these works, the Laplacian calculation with neural networks is indispensable. Our proposed Forward Laplacian offers advantages for all the aforementioned methods and can be integrated with all existing NN-VMC packages \cite{deepqmc,netket3:2022,FermiNet2020github}.


Although our proposed approach yields relative energies that align closely with the gold standard or experimental results in most cases, there are instances where discrepancies persist between our results and ground truth. Our hypothesis for this inconsistency is that existing NN-VMC methods inadequately incorporate all critical chemical and physical knowledge. For instance, the error cancellation property, which is crucial for obtaining accurate relative energies in gold-standard methods like CCSD(T), is not inherently encoded in the NN-VMC ansatz or prevailing training strategies. We believe encoding proper chemical and physical knowledge into neural networks will be vital in advancing NN-VMC techniques in the future.


Our principal objective is to tackle the computational bottleneck associated with the calculation of the Laplacian in the NN-VMC methods. However, the prospective utilization of our Forward Laplacian approach could extend to diverse scenarios outside the realm of quantum mechanics. As shown in the Method section, the Forward Laplacian method can replace the previous method with improved efficiency when calculating Laplacian associated with any neural network. Therefore, it can accelerate the training of other neural network-based solvers of partial differential equations \cite{wang20222,raissi2019physics,he2023learning}, such as heat diffusion equations and Navier-Stokes equations. 

\section{Methods}\label{sec: Methods}

\subsection{Wavefunction Optimization}\label{sec: optimization}

\newcommand{\B}[1]{\mathbf{#1}}
\newcommand{\norm}[1]{\left| #1 \right|}


For a molecular system with $N$ electrons and $M$ nuclei, we consider the time-independent Schr\"{o}dinger equation under the Born-Oppenheimer approximation:
\begin{align}
    \hat{H} & \psi\left(\mathbf{x}_1, \ldots, \mathbf{x}_N\right)=E \psi\left(\mathbf{x}_1, \ldots, \mathbf{x}_N\right),\\
    \hat{H}= & -\frac{1}{2} \sum_i \Delta_i+\sum_{i>j} \frac{1}{\norm{\mathbf{r}_i-\mathbf{r}_j}} \\ 
        & -\sum_{i I} \frac{Z_I}{\left|\mathbf{r}_i-\mathbf{R}_I\right|}+\sum_{I>J} \frac{Z_I Z_J}{\left|\mathbf{R}_I-\mathbf{R}_J\right|},
\end{align}
where $\mathbf{x}_i=\{\mathbf{r}_i,\sigma_i\}$ is the coordinate of electron $i$. $\mathbf{r}_i\in\mathbb{R}^3$ is the position of the $i$-th electron and $\sigma_i\in\{1,-1\}$ is the spin coordinate. $Z_I$ is the charge of the $I$-th nucleus and $\mathbf{R}_I$ is the corresponding position. $\Delta_i$ is the Laplacian w.r.t. $\mathbf{r}_i$. For ease of reference, we denote $(\rvx_1,\rvx_2,...,\rvx_N)$ as $\rvx$. 

The wavefunction of a multi-electron system should be antisymmetric under the exchange of any electron's coordinates of the same spin, i.e.,
\begin{equation}
    \psi(\cdots,\rvx_i,\cdots,\rvx_j)=-\psi(\cdots,\rvx_j,\cdots,\rvx_i).
\end{equation}
Following previous methods, we leverage the form of Slater-Jastrow-Backflow ansatz to satisfy this condition:
\begin{equation}
\begin{aligned}
    \psi_\theta(\rvx)&=e^{J_\theta(\rvx)} \sum_{k=1}^{K}\operatorname{det}\left[\boldsymbol{\Phi}_\theta^k(\rvx)\right],
\end{aligned}
\end{equation}
where $J_{\theta}(\rvx)$ is the Jastrow factor and $\boldsymbol{\Phi}^k_{\theta}(\rvx)$ is a permutation-equivariant neural network. The output of this network is used as the input for the $k$-th Slater determinant. 

Following \cite{ferminet}, the training of NN-VMC is divided into two phases. The first is the pre-training phase in which $\boldsymbol{\Phi}_\theta^k(\rvx)$ is updated to match the Hartree-Fock orbitals. The second is the VMC training phase where we optimize the parameters to minimize the total energy of the wavefunction defined below. 
\begin{equation}
\label{eq: VMC loss}
    \begin{aligned}
        \mathcal{L}_{\theta} & =\frac{\langle\psi_{{\theta}}|\hat{H}| \psi_{{\theta}}\rangle}{\langle\psi_{{\theta}}| \psi_{{\theta}}\rangle}=\frac{\int \psi^*_\theta(\rvx)\hat{H}\psi_\theta(\rvx)d\rvx}{\int \psi^*_\theta(\rvx)\psi_\theta(\rvx)d\rvx}\\ & =\int p(\rvx)E_L(\rvx)d\rvx.
    \end{aligned}
\end{equation}
$E_L(\rvx)=\psi^{-1}_\theta(\rvx)\hat{H}\psi_\theta(\rvx)$ is known as the local energy, and $p(\boldsymbol{\rvx})=\frac{\psi^*_\theta(\mathbf{x})\psi_\theta(\mathbf{x})}{\langle\psi_\theta|\psi_\theta\rangle}$.
We use gradient descent to optimize the parameters, where the unbiased estimate of $\theta$'s gradient is given by:
\begin{align}
\label{eq: VMC gradient}
&\nabla_\theta \mathcal{L}_{\theta} = 
2 \mathbb{E}_{p(\mathbf{x})}[(E_L-\mathbb{E}_{p(\mathbf{x})}[E_L]) \nabla_\theta \log |\psi_\theta|].
\end{align}

The loss function and the gradient are estimated using the Metropolis-Hastings algorithm, one of the most popular Markov Chain Monte Carlo (MCMC) methods. To make the initial electron distribution closer to the target distribution, we assign each electron to a nucleus according to the Mulliken population analysis \cite{mulliken1955electronic} results within the Hartree-Fock method. Initial electron positions are then sampled from Gaussian distributions centered around the corresponding atom. The convergence of the wavefunction to the ground state is achieved through iterative sampling and updates of parameters during training. For evaluation, we follow \cite{ferminet} and run MCMC steps without updating the network parameters to sample batches of workers and compute their local energy. The total energy and associated standard error are determined by reblocking analysis.

In practice, the accuracy of a learned model can be significantly influenced by the quality of its pre-training phase. The Hartree-Fock solution may be inaccurate when the electron structure is complex, such as in transition metals, which may result in an inaccurate initialization of the neural network during a prolonged pre-training phase. To mitigate this issue, we use a short pre-training phase alongside a moderate basis set 
for the Hatree-Fock method. This strategy results in improved performance on complex systems like \ce{CCl4}. All training hyperparameters are listed in \sptable{2 and 3}.



\subsection{\fl Method for General Neural Networks}\label{subsec: method forward lap}

In this subsection, we provide a detailed explanation of how the \fl method accelerates the calculation of Laplacian for general neural networks. The computation graph serves as the descriptive language of deep learning models across various deep learning toolkits, including PyTorch \cite{paszke2017automatic}, TensorFlow \cite{abadi2016tensorflow}, and Jax \cite{jax2018github}. Therefore, we can narrow our focus to investigating the performance of the proposed method in a specific computation graph.

In a computation graph $\mathcal{G}$, the edges represent function arguments, and nodes represent operations or variables. Note that $\mathcal{G}$ is always a directed acyclic graph. We use $\{{\rvv}^{-1},...{\rvv}^{-N}\}$ to represent the external nodes of $\mathcal{G}$ (i.e., the input of the neural network), and use $\{\rvv^0,...\rvv^{M}\}$ to represent the internal nodes, sorted in topological orders. $\rvv^{M}$ serves as the network output $\phi$. We use the abbreviation $i\to j$ if there is a directed edge from $\rvv^i$ to $\rvv^j$ in  $\mathcal{G}$. Furthermore, we denote operations as $F$, e.g., $\rvv^j=F_j(\{\rvv^i:i\to j\})$ for all $j\geq 0$. 

We first show that the \fl method can be applied to general neural networks, such as FermiNet, Psiformer, and \name. Similar to Section \ref{2.1}, along the computation graph, we compute the tuple $(v_j, \nabla v_j, \Delta v_j)$ for each node. To be specific, the tuple associated with node $j$ is derived through chain rules\cite{botev2020gauss}:

\begin{align}
        \rvv^j &= F_j(\{\rvv^i:i\to j\})\label{eq:fl_method1} \\
        \nabla_{ } \rvv^{j} &= \sum\limits_{i:i\to j}\pab{F_{j}}{\rvv^i}\nabla_{ } \rvv^{i} \label{eq:fl_method2}\\
        \Delta \rvv^{j} &= \sum\limits_{\substack{i,l\\i\to j\ l\to j}}\ppabc{F_{j}}{\rvv^{i}}{\rvv^{l}}\nabla \rvv^{i}\cdot \nabla \rvv^{l} + \sum\limits_{i:i\to j} \pab{F_{j}}{\rvv^{i}}\Delta \rvv^{i}\label{eq:fl_method3}
\end{align}
Note that eqs. (\ref{eq:fl_method1}) - (\ref{eq:fl_method3}) only rely on tuples associated with parent nodes of $j$. Therefore, we can sequentially apply eqs. (\ref{eq:fl_method1}) - (\ref{eq:fl_method3}) to nodes of any computational graph in topological orders, and finally get the Laplacian in the output node.

We then analyze the speed-up rate of the \fl method compared to the previous method. The previous method obtains $\Delta\phi(\rvx)$ through computing the Hessian matrix. It first performs forward propagation to obtain the value of each variable $\rvv^j$. Next, a standard backward process is employed by creating a new computation graph $\hat{\mathcal{G}}$ where node $\hat{\rvv}^i\in \hat{\mathcal{G}}$ represents the operation to calculate $\pab{\phi}{\rvv^i}$, $i=M,...,-N$. The associated computations are
\begin{equation}
\label{eq: backward}
    \pab{\phi}{\rvv^i} = \sum\limits_{j:i\to j} \pab{F_j}{\rvv^i} \pab{\phi}{\rvv^j},\ i=M-1,...,-N.
\end{equation}

The Hessian matrix is then obtained through the following forward-mode Jacobian calculation along $\mathcal{G}$ and $\hat{\mathcal{G}}$, respectively:
\begin{equation}
\label{eq: forward grad}
    \begin{aligned}
    \nabla_{ } \rvv^{i} &= \sum\limits_{j:j\to i}\pab{F_{i}}{\rvv^j}\nabla_{ } \rvv^{j} \quad 
    \end{aligned}
\end{equation}
\begin{equation}
\label{eq: hessian second}
    \begin{aligned}
    \nabla_{ } \pab{\phi}{\rvv^{i}}  =& \sum\limits_{\substack{j,l\\ i\to j\\ l\to j}} \ppabc{F_{j}}{\rvv^{l}}{\rvv^{i}}\pab{\phi}{\rvv^{j}}\nabla_{ } \rvv^{l} + \sum_{j:i\to j}\pab{F_{j}}{\rvv^{i}}\nabla_{ }\pab{\phi}{\rvv^{j}}, \\
\end{aligned}
\end{equation}
where we always use $\nabla$ to denote $\nabla_\rvx$ for simplicity.

The bottleneck of Hessian computation comes from eq. (\ref{eq: forward grad}) and eq. (\ref{eq: hessian second}). 
eq. (\ref{eq: forward grad}) takes $N|E|$ floating point operations (FLOPs) if we only count multiplications, where $E$ denotes the set of edges in $\mathcal{G}$. For eq. (\ref{eq: hessian second}), the second term also takes $N|E|$ FLOPs. To calculate the computational cost of the first term in eq. (\ref{eq: hessian second}), we first introduce two notations, $T$ and $R$, which are both sets of ordered tuples:
\begin{equation}
    \begin{aligned}
        T &= \{(i,l,j)| i\to j ,\ l \to j,\ \ppabc{F_j}{\rvv^i}{\rvv^l}\neq 0\},\\
        R &= \{(i,l)| \exists j\ s.t.\ (i, l, j)\in T\}.
    \end{aligned}
\end{equation}
The previous method sums over $j$ first to obtain $\sum\limits_{j:i\to j,l\to j} \ppabc{F_{j}}{\rvv^{l}}{\rvv^{i}}\pab{\phi}{\rvv^{j}}$ for all $(i,l)\in R$, and then sums over $l$. It can be shown that the total computation cost of this term is $0.5|T|+N|R|$ by exploiting the symmetry of the Hessian matrix. Thus, the total FLOPs for the previous method is about $N(|R|+2|E|)+0.5|T|$.


For the proposed \fl method, we perform forward propagation along $\mathcal{G}$ to obtain $\phi(\rvx)$, $\nabla\phi(\rvx)$ and $\Delta\phi(\rvx)$. The second term $\nabla\phi(\rvx)$ is calculated along $\mathcal{G}$ according to eq. (\ref{eq:fl_method2}). The third term, i.e., the Laplacian, is calculated along $\mathcal{G}$ according to eq. (\ref{eq:fl_method3}).

The computational cost of the \fl method is dominated by eq. (\ref{eq:fl_method2}) and eq. (\ref{eq:fl_method3}). As previously discussed, eq. (\ref{eq:fl_method2}) takes $N|E|$ FLOPs.
For the first term in eq. (\ref{eq:fl_method3}), we decompose its calculation into two steps. First, we compute $\{\nabla \rvv^{j}\cdot \nabla \rvv^{l}\}_{j\leq l,\ (j, l)\in R}$, which takes $0.5N|R|$ FLOPs in total. Next, following the topological order of $\mathcal{G}$, we sum over $j\leq l$ for each $i$, deriving the first term in $\Delta\rvv^i$. By leveraging the symmetry of the Hessian matrix and Gram matrix, we reduce this computation by a factor of 2, which is $0.5|T|$ FLOPs. The computational cost of the second term is negligible compared with the first term since $\Delta \rvv^{j}$ is a scalar. Summing the computational cost of all the terms, we have that the \fl method uses $0.5N(|R|+2|E|)+0.5|T|$. In practice, a large percentage of operations are linear transformations, and for any linear operation $F_j$, $\ppabc{F_j}{\rvv^i}{\rvv^l}=0$ for any $i\to j,\ l\to j$. This means the value $|T|$ is much smaller than $N|R|$ and $N|E|$. Thus, our method is about two times faster than the previous method for computing Laplacian.





We implement the \fl method in \texttt{Jax}\cite{jax2018github}. We overload a large number of frequently used operations such that if the input of $f(\rvx)$ is a Laplacian tuple $\left(\rvx,\nabla \rvx,\Delta \rvx\right)$, the output will automatically become another Laplacian tuple $\left(f(\rvx),\nabla f(\rvx), \Delta f(\rvx)\right)$. This process is implemented via \texttt{Python} decorator, which can make our implementation compatible with existing operations. To achieve further acceleration, we manually define the forward propagation rules for a wide range of operations used in NN-VMC methods and optimize their efficiency in computing the Laplacian tuple. Specifically, we mainly optimize two classes of operations, namely \verb|element-wise operations|, \verb|linear operations|, and several frequently used non-linear operations. 

Element-wise operations apply a scalar function to each element of the input array and output an array with the same shape. Many operations in neural networks are element-wise, such as activation and exponential functions. The Jacobian matrix is always diagonal for these operations, and there is no need to calculate non-diagonal elements. We implement a customized framework for element-wise functions to avoid unnecessary calculations, thus speeding up the Laplacian computation. An operation $f$ is linear if $f(\rvx+ \rvy) = f(\rvx) + f(\rvy)$. Here $\rvx, \rvy$ are arrays. For example, \verb|reshape, concatenate, sum| are all linear. It can be proved that for linear operations, we have $\nabla f(\rvx) = f(\nabla \rvx)$. Based on this, we can compute $\nabla f(\rvx), \Delta f(\rvx)$ by directly ``applying'' $f$ on $\nabla \rvx, \Delta \rvx$, thus avoiding second-order calculations. For operations that are not mentioned above, but are commonly used such as \verb|matmul, slogdet, softmax|, we manually customize the computation rules case-by-case as they significantly differ from one another.

\subsection{\name Ansatz} \label{sec: ansatz}

To better leverage the benefits of the \fl method, we develop a neural network architecture, called \name. This architecture facilitates a significant faster Laplacian calculation \cite{martens2012estimating} thus improves the training speed further without sacrificing any accuracy compared to the previous state-of-the-art approach \cite{glehn2023a}. The overall architecture of \name is illustrated in Fig. \ref{fig:network}.


We use the relative position of electron-nuclei pairs $\rvr_{i}-\mathbf{R}_{I}$ and the electron spins $\sigma_i$ as the input features. Following \cite{glehn2023a}, a norm-rescaled function $\text{Rescale}: \mathbb R ^{3}\to \mathbb R ^{4}$ is applied to each electron-nuclei pair:
\begin{equation}
\begin{aligned}
    &{\bf e}_{iI} = \text{Rescale}(\rvr_i-{\mathbf{R}}_I),
\end{aligned}
\end{equation}
where each output dimension of $\text{Rescale}(\rvz)$ is given by:
\begin{equation}
\begin{aligned}
    &\text{Rescale}_1(\rvz) =\ln (1+|\rvz|)\hat \rvz_1\\
    &\text{Rescale}_2(\rvz) =\ln (1+|\rvz|)\hat \rvz_2\\
    &\text{Rescale}_3(\rvz) =\ln (1+|\rvz|)\hat \rvz_3\\
    &\text{Rescale}_4(\rvz) = \ln (1+|\rvz|).
\end{aligned}
\end{equation}
$|\cdot|$ refers to the $L_2$ norm and $\hat \rvz = \rvz/|\rvz|$ is the directional vector of $\rvz$, the subscripts refer to the index of coordinates. Then, $\{e_{iI}\}_{I=1}^{M}$ is concatenated with $\sigma_i$ to form the feature set for electron $i$:
\begin{equation}
    {\bf d}_{i}=\text{Concat}(\{{\bf e}_{iI}\}_{I=1}^{M}, \sigma_i).
\end{equation}
${\bf d}_i$ is then projected via a linear layer into two terms $\B{h}^0_i,\B{g}^0_i$, which will be further fed into the \name. We denote $\B{h}^0=(\B{h}^0_1,\cdots,\B{h}^0_i,\cdots,\B{h}^0_N)$ and $\B{g}^0=(\B{g}^0_1,\cdots,\B{g}^0_i,\cdots,\B{g}^0_N)$ for ease of reference.

The main building block of \name is the stacked Sparse Derivative Attention (SDA) module. The $l$-th SDA module takes $\B{h}^l,\B{g}^l$ as input and outputs $\B{h}^{l+1},\B{g}^{l+1}$ through two streams. 

$\B{g}^l$ is updated by a stream called the individual stream, where each $\B{g}^{l}_i$ is updated as below
\begin{align}
\B{g}^{l+1}_i&=\B{g}^{l}_i+\mathrm{MLP}^l_{\mathrm{indiv}}\left(\B{g}^l_i\right).
\end{align}
In the individual stream, each electron $i$ independently updates its own features using MLPs with parameter sharing. Since there is no interaction between different electrons in the stream, the derivative matrix $\nabla_{\rvr}\B{g}^{l}$ is sparse, i.e., $\nabla_{\rvr_{i'}}\B{g}_i^{l}=0$ when $i\neq i'$. As shown in Section \ref{sec: sparsity}, this sparsity leads to acceleration with the \fl method.

$\rvh^{l}$ is updated by the other stream called attentive stream, where the features of different electrons interact with each other through attention. In the attention, we use individual-stream-produced $\B{g}^l$ as queries and keys, and use attention-stream-produced $\B{h}^l$ as Values with different linear projection layers.
\begin{align}
    {\bf q}_i =  W_Q^{l} \B{g}_i^l, {\bf k}_i =  W_K^{l} \B{g}_i^l, {\bf v}_i =  W_V^{l} \B{h}_i^l,
\end{align}
Matrices $W_Q^{l}$, $W_K^{l}$ and $W_V^{l}$ are learnable parameters of the $l$-th block. We use normalized dot products to calculate weights assigned to different value terms and the update of $\rvh^{l}$ is given by
\begin{align}
    \bar{\B{h}}^{l+1}_i&=\B{h}^l_i+ \sum_j \alpha_{ij} \rvv_j \label{eq: attention}\\    \B{h}^{l+1}_i&=\bar{\B{h}}^{l+1}_i+\mathrm{MLP}^l_{\mathrm{attn}}\left(\bar{\B{h}}^{l+1}_i\right),
\end{align}
where
\begin{align}
    &\alpha_{ij} = \frac{S_{ij}}{\sum_{j'} S_{ij'}}, S_{ij} = \exp({\bf q}_i^{\top}{\bf k}_j)\label{eq: alpha}
\end{align}

Note that queries and keys used in the attentive stream are independently updated in the individual stream. Therefore the $S_{ij}$ also exhibit a certain level of derivative sparsity. As a result, this design facilitates efficient computation through our Forward Laplacian method.

After stacking $L$ layers of Sparse Derivative Attention blocks, we use the output of attentive stream $\B{h}^L_i$ to construct the wavefunction. Our wavefunction follows the Slater-Jastrow-Backflow ansatz \cite{10.1063/1.2743972}:
\begin{equation}
\begin{aligned}
    \psi_\theta(\rvx)&=e^{J_\theta(\rvx)} \sum_{k=1}^{K}\operatorname{det}\left[\boldsymbol{\Phi}_\theta^k(\rvx)\right]. \label{eq: SJ}
\end{aligned}
\end{equation}
To construct $\boldsymbol{\Phi}_\theta^k$, we first project $\B{h}^L_i$ to $\mathbb R^{N}$ using a spin-dependent linear transformation:
\begin{align}
    \B{o}^k_i=\B{W}_{\sigma_i}^k \B{h}_i^{L} + \B{b}_{\sigma_i}^k,
\end{align}
To ensure the wavefunction approaches zero when the electron moves away from all nuclei, we use the envelope function and multiply it with $\B{o}^k_i$ to obtain $\boldsymbol{\Phi}_\theta^k(\rvx)$:
\begin{align}
 \text{env}_j(\rvx_i)&=\sum_{I} {\pi^{\sigma_i}_{Ij} e^{-|\xi^{\sigma_i}_{Ij}| |\mathbf{r}_i - \mathbf{R}_I|}},\\
    [\boldsymbol{\Phi}_\theta^k(\rvx)]_{ji} &= \text{env}_j(\rvx_i)[\B{o}^k_i]_j , 
\end{align}
where $\pi^{\sigma_i}_{Ij}$ and $\xi^{\sigma_i}_{Ij}$ are learnable parameters. In practice, we observed some determinants $\text{det}[\boldsymbol{\Phi}_\theta^k(\rvx)]$ degenerate and we remove those determinants during training. We use a simple form for the Jastrow factor in eq. (\ref{eq: SJ}) to satisfy the electron-electron Cusp condition \cite{PhysRevB.70.235119}: 
\begin{equation}
\begin{aligned}
        J_{\theta}(\rvx) &= \sum_{i<j;\sigma_i=\sigma_j}-\frac{1}{4}\frac{\alpha^2_{\text{par}}}{\alpha_{\text{par}}+|\rvr_i-\rvr_j|} \\
        & \quad + \sum_{i<j;\sigma_i\neq\sigma_j}-\frac{1}{2}\frac{\alpha^2_{\text{anti}}}{\alpha_{\text{anti}}+|\rvr_i-\rvr_j|},
\end{aligned}
\end{equation}
where $\alpha_{\text{par}}$ and $\alpha_{\text{anti}}$ are learnable parameters. Combining with $\text{det}[\boldsymbol{\Phi}_\theta^k(\rvx)]$, the wavefunction is finally computed through eq. (\ref{eq: SJ}). As discussed above, the \name incorporates more derivative sparsity in many operations related to $\rvg^l_i$, ${\bf q}_i$, ${\bf k}_i$ and $S_{ij}$, leading to a significantly speed-up rate compared to previous architectures.

\backmatter





\bmhead{Acknowledgments}

We thank Hang Li and ByteDance Research for support and inspiration. We thank David Pfau for his valuable feedback. 
We thank Bohang Zhang and Haiyang Wang for their helpful suggestion and discussion. Liwei Wang is supported by National Key R\&D Program of China (2022ZD0114900) and National Science Foundation of China (NSFC62276005).
J.C. acknowledges the National Natural Science Foundation of China for support under Grant No. 92165101. 

\bibliography{sn-bibliography}




\end{document}


\title{Supplementary information of ``Forward Laplacian: A New Computational Framework for Neural Network-based Variational Monte Carlo''}

\maketitle
\section{Hyperparameters}
\sptable{1} gives the network architecture hyperparameters for each NN-VMC ansatz. The default hyperparameters for all calculations reported in the main text are mainly listed in \sptable{2}. The variations in hyperparameters are shown in \sptable{3}. 

\begin{table*}[htbp]
    \centering
    \caption{Architecture hyperparameter for different NN-VMC models 
    }
    \begin{tabular}{cccc}
    \toprule\midrule
    Parameter & FermiNet & Psiformer & LapNet \\
    \midrule
    Determinants & 16 & 16 & 16 \\
    Network layers & 4 & 4 & 4 \\
    Attention heads & \textendash & 4 & 4  \\
    Attention dimension & \textendash & 64 & 64 \\
    MLP hidden dimension & (256, 32) & 256 & 256 \\
    \midrule\bottomrule
    \end{tabular}
    \label{table:nets}
\end{table*}

\begin{table*}[htbp]
    \centering
    \caption{Default hyperparameters}
    \begin{tabular}{ccc}
    \toprule\midrule
        & Parameter & Value \\
    \midrule
    \multirow{6}{*}{Training}& Optimizer & KFAC \\
    & Iterations & 2e5 \\
    & Learning rate at iteration $t$ & ${lr}_0/(1+\frac{t}{t_0})$ \\
    & Initial learning rate ${lr}_0$ & 0.05\\
    & Learning rate decay $t_0$ & 1e4\\
    & Local energy clipping & 5.0 \\
    \midrule
    \multirow{2}{*}{Pretraining} & Optimizer & LAMB \\
    & Learning rate &1e-3\\
    \midrule
    \multirow{2}{*}{MCMC} & Decorrelation steps & 30 \\
    & Proposal standard deviation & 0.02 \\
    \midrule
    \multirow{4}{*}{KFAC} & Norm constraint & 1e-3 \\
    & Damping & 1e-3 \\
    & Momentum & 0 \\
    & Covariance moving average decay & 0.95 \\
    \midrule\bottomrule
    \end{tabular}
    \label{table:default}
\end{table*}


\begin{table*}[htbp]
    \centering
    \addtolength{\leftskip} {-2cm}
    \addtolength{\rightskip}{-2cm}    \caption{Variations in hyperparameters between calculations}
    \begin{tabular}{cccccc}
    \toprule\midrule
        ~ & Number of  & Pretraining  & MCMC & Pretraining  & Batch  \\ 
        & electrons $n$ & iterations& blocks& basis set& size\\\midrule
        \multirow{3}{*}{Molecules} & $n<42$ & 5e3 & 1 & \multirow{3}{*}{cc-pVDZ} & \multirow{3}{*}{4096} \\ 
        ~ & $42\leq n<84$ & 2e4 & 2 & ~ & ~ \\ 
        ~ & $n\geq84$ & 5e4 & 4 & ~ & ~ \\ 
        \midrule
        Ionization potential & any & 5e3 & 1 & cc-pVTZ & 8192 \\\midrule
        Barrier height & any & 2e4 & 1 & aug-cc-pVDZ & 4096 \\ \midrule
         \multirow{3}{*}{Interaction energy} & $n<42$ & 5e3 & 1 &  \multirow{3}{*}{aug-cc-pVDZ} & \multirow{3}{*}{4096} \\ 
         & $42\leq n<84$ & 2e4 & 2 &  & \\ 
         & $n\geq 84$ & 5e4 & 4 &  & \\ \midrule\bottomrule
    \end{tabular}
    \label{table:variations}
\end{table*}

\newpage
\section{Systems}

The configurations of molecules studied in Fig. 3a  can be found in \cite{ferminet,glehn2023a}. The reactions we studied in Fig. 4b are listed in \sptable{4} and the corresponding configurations can be found in \cite{C7CP04913G}. The non-covalent systems we studied in Fig. 4c are listed in \sptable{5} and the corresponding configurations can be found in \cite{jurevcka2006benchmark}. 

\begin{table*}[htbp]
    \centering
    \caption{Reactions in Fig. 4b 
    }
    \begin{tabular}{lll}
    \toprule\midrule
     & Reaction & Type \\
    \midrule
    1&\ce{H + N2O -> OH + N2}&\multirow{3}{*}{Heavy-atom transfer reaction}\\
    2&1 reverse&\\
    3&\ce{H + FH -> HF + H}&\\
    \midrule
    4&\ce{F- ... CH3F -> FCH3 ... F-}&\multirow{2}{*}{Nucleophilic substitution reaction}\\
    5&\ce{Cl- ... CH3Cl -> ClCH3 ... Cl-}&\\
    \midrule
    6&\ce{H + C2H4 -> C2H5}&\multirow{5}{*}{Unimolecular and association reaction}\\
    7&\ce{C2H5 -> H + C2H4}&\\
    8&\ce{HCN -> HNC}&\\
    9&8 reverse&\\
    10&\ce{\textit{s-trans\ cis-}C5H8 -> \textit{s-trans\ cis-}C5H8}&\\
    \midrule
    11&\ce{NH2 + CH3 -> NH + CH4}&\multirow{4}{*}{Hydrogen transfer reaction}\\
    12&11 reverse&\\
    13&\ce{NH2 + C2H5 -> NH + C2H6}&\\
    14&13 reverse&\\
    \midrule\bottomrule
    \end{tabular}
    \label{table:systems bh76}
\end{table*}

\begin{table*}[htbp]
    \centering
    \caption{Systems in Fig. 4c
    }
    \begin{tabular}{lll}
    \toprule\midrule
     & system & Type \\
    \midrule
    1&Water dimer&\multirow{4}{*}{Hydrogen-bonding interaction system}\\
    2&Formic acid dimer&\\
    3&Formamide dimer&\\
    4&Uracil dimer h-bonded&\\
    \midrule
    5&Methane dimer&\multirow{3}{*}{Dispersion interaction system}\\
    6&Ethene dimer&\\
    7&Uracil dimer stack&\\
    \midrule
    8&Ethene ethyne complex&\multirow{4}{*}{Mixed electrostatic dispersion interaction system}\\
    9&Benzene water complex&\\
    10&Benzene dimer T-shaped&\\
    11&Phenol dimer&\\
    \midrule\bottomrule
    \end{tabular}
    \label{table:systems s22}
\end{table*}
\section{Energy results}
This note provides the energy results plotted in Fig. 3a and Fig. 4. 
\begin{table}[!ht]
    \centering
    \caption{Molecules energy in Fig. 3a. The results of FermiNet and Psiformer are from \cite{glehn2023a}.}
    \begin{tabular}{c r r r r}
    \toprule\midrule
        System & HF, aug-cc-pVDZ& Psiformer\cite{glehn2023a}& FermiNet\cite{glehn2023a} & LapNet \\ \midrule
        \ce{LiH} & -7.98416 & -8.070528(5) & -8.07050(1) & -8.070523(4) \\ 
        \ce{Li2} & -14.86984 & -14.99486(1) & -14.99480(2) & -14.99485(1) \\ 
        \ce{NH3} & -56.20514 & -56.56367(2) & -56.56347(4) & -56.56359(2) \\
        \ce{CH4} & -40.19963 & -40.51454(2) & -40.51430(3) & -40.51445(2) \\ 
        \ce{CO} & -112.75173 & -113.32416(4) & -113.32354(7) & -113.32417(4) \\ 
        \ce{N2} & -108.96104 & -109.54137(4) & -109.54046(6) & -109.54128(4) \\ 
        \ce{C2H4} & -78.04351 & -78.58762(3) & -78.58604(5) & -78.58721(4) \\ 
        methylamine & -95.23044 & -95.86050(4) & -95.85917(6) & -95.86025(3) \\ 
        \ce{O3} & -224.27901 & -225.43061(9) & -225.4226(2) & -225.4293(1) \\ 
        ethanol & -154.10515 & -155.04656(7) & -155.0419(1) & -155.04563(6) \\ 
        bicyclobutane & -154.88814 & -155.94619(8) & -155.9388(1) & -155.94528(4) \\ \midrule
        benzene & -230.7280 & -232.2400(1) & -232.2205(2) & -232.2389(1) \\ 
        toluene & -269.7684 & -271.5538(1) & -271.5274(2) & -271.5524(1) \\ 
        naphthalene & -383.3934 & -385.8685(2) & -385.8147(4) & -385.8655(2) \\ 
        \ce{CCl4} & -1875.8507 & -1878.804(1) & -1878.684(1) & -1878.839(1) \\ 
        \midrule
        benzene dimer & -461.4555 & -464.4667(2) & -464.3770(5) & -464.4681(1) \\ \midrule\bottomrule
    \end{tabular}
    \label{table:energy 3a}
\end{table}
\clearpage
\begin{table}[!ht]
    \centering
    \caption{Ionization potential of 5 transition metals in Fig. 4a}
    \begin{tabular}{c r r r r}
    \toprule\midrule
        System & Atom & Ion & Ionization potential & Experiments\cite{10.1063/1.2335444} \\ \midrule
        V & -943.8773(1) & -943.6412(1) & 0.2361(2) & 0.23733 \\ 
        Fe & -1263.6586(2) & -1263.3743(2) & 0.2843(3) & 0.2871 \\ 
        Co & -1382.7192(1) & -1382.4487(1) & 0.2705(2) & 0.27137 \\ 
        Ni & -1508.2694(2) & -1508.0117(2) & 0.2577(2) & 0.25864 \\ 
        Zn & -1779.4364(2) & -1779.1017(3) & 0.3347(4) & 0.3383 \\ \midrule\bottomrule
    \end{tabular}
\end{table}

\begin{table}[!ht]
    \centering
    \caption{Barrier heights of 14 reactions in Fig. 4b}
    \begin{tabular}{c r r r r}
    \toprule
    \midrule
        ~ & Reactants  & Transition state & Barrier height& W2-F12\cite{C7CP04913G} \\ \midrule
        1 & -185.17156(5) & -185.14391(5) & 0.02765(7) & 0.02821 \\ 
        2 & -185.27710(4) & -185.14391(5) & 0.13319(6) & 0.13163 \\ 
        3 & -100.95854(3) & -100.89248(2) & 0.06606(4) & 0.06709 \\ \midrule
        4 & -239.62929(5) & -239.60805(7) & 0.02124(9) & 0.02135 \\ 
        5 & -960.3632(5) & -960.3388(5) & 0.0244(7) & 0.02151 \\ \midrule
        6 & -79.08694(2) & -79.08432(2) & 0.00262(3) & 0.00319 \\ 
        7 & -79.15115(2) & -79.08432(2) & 0.06683(3) & 0.06693 \\
        8 & -93.43241(1) & -93.35566(2) & 0.07675(3) & 0.07665 \\
        9 & -93.40824(2) & -93.35566(2) & 0.05258(3) & 0.05259 \\
        10 & -195.30143(7) & -195.23616(7) & 0.0653(1) & 0.06327 \\ \midrule
        11 & -95.71460(2)& -95.70018(3) & 0.01442(3) & 0.01418 \\
        12 & -95.73547(2)& -95.70018(3) & 0.03529(3) & 0.03506 \\ 
        13 & -135.02977(3) & -135.01383(3) & 0.01594(5) & 0.01562 \\ 
        14 & -135.04518(4) & -135.01383(3) & 0.03135(5)& 0.03092 \\ \midrule\bottomrule
    \end{tabular}
\end{table}

\begin{table}[!ht]
    \centering
    \caption{Interaction energies of 11 systems in Fig. 4c}
    \begin{tabular}{c c r r r r }
    \toprule\midrule
        &System & Equilibrium & Dissociated & Interaction  & CCSD(T)/CBS\cite{10.1063/1.3659142}  \\ \midrule
        1&Water dimer & -152.88309(6) & -152.87584(3) & 0.00725(8) & 0.00795 \\ 
        2&Formic acid dimer & -379.5638(1) & -379.5374(1) & 0.0264(2) & 0.02988\\ 
        3&Formamide dimer & -339.8195(1) & -339.79692(8) & 0.0226(1) & 0.02560 \\ 
        4&Uracil dimer h-bonded & -829.6199(3) & -829.5829(3) & 0.0370(4) & 0.03289 \\ 
        \midrule
        5&Methane dimer & -81.02888(8) & -81.02861(2) & 0.0003(1) & 0.00084\\ 
        6&Ethene dimer & -157.17415(7) & -157.17258(5) & 0.0016(1) & 0.00235 \\ 
        7&Uracil dimer stack & -829.5946(3) & -829.5829(3) & 0.0117(4) & 0.01563 \\ \midrule
        8&Ethene-ethyne complex & -155.92198(6) & -155.92068(5) & 0.00130(8) & 0.00238 \\ 
        9&Benzene-water complex & -308.6756(1) & -308.6727(1) & 0.0029(2) & 0.00522 \\
        10&Benzene dimer T-shape & -464.4681(1) & -464.4655(1) & 0.0026(1) & 0.00433\\ 
        11&Phenol dimer & -614.9097(2) & -614.9017(2) & 0.0080(3) & 0.01131 \\ \midrule\bottomrule
    \end{tabular}
\end{table}

\renewcommand{\refname}{Supplementary References}
\bibliography{sn-bibliography}